\documentstyle[12pt]{article}
\renewcommand{\topmargin}{-1.5cm}  

\def\mk{m_K}
\def\mp{m_\pi}
\def\SS{\scriptscriptstyle}

\title{
\vspace*{-1.3cm}
\begin{flushright}
{\normalsize DO--TH 99/16\\[-10pt]
LNF-99/020(P)}
\end{flushright}
\vspace{0.8cm}
{\Large \bf \boldmath $\varepsilon'/\varepsilon$ \unboldmath in the 
\boldmath $1/N_c$ \unboldmath Expansion}\footnote{Talk presented by 
T.\ Hambye at the 1999 Chicago Conference on Kaon Physics (KAON~99), 
Chicago, IL, 21-26 Jun 1999.}
\vspace*{0.5cm}
}
\author{
T.~Hambye$^{a\,}$ and P.H. Soldan$^{b\,}$
\\[0.5cm]
\small $a$: {\it INFN - Laboratori Nazionali di Frascati,
I-00044 Frascati, Italy}\\ 
\small $b$: 
{\it Institut f\"ur Physik, Universit\"at Dortmund 
D-44221 Dortmund, Germany}\\[2.4cm] 
}
\date{}
\begin{document}
\maketitle
\thispagestyle{empty}
\vspace*{-2.9cm}
\begin{abstract}
We present a new analysis of the ratio $\varepsilon'/\varepsilon$ which 
measures the direct CP violation in $K\rightarrow\pi\pi$ decays. We use the 
$1/N_c$ expansion within the framework of the effective chiral lagrangian 
for pseudoscalar mesons. From general counting arguments we show that 
the matrix element of the operator $Q_6$ is not protected from possible 
large $1/N_c$ corrections beyond its large-$N_c$ or VSA value. Calculating 
the $1/N_c$ corrections, we explicitly find that they are large and positive. 
Our result indicates that a $\Delta I=1/2$ enhancement is operative for 
$Q_6$ similar to the one of $Q_1$ and $Q_2$ which dominate the CP 
conserving amplitude. This enhances $\varepsilon'/\varepsilon$ and can 
bring the standard model prediction close to the measured value for 
central values of the parameters. 
\end{abstract}
\vspace*{0.5cm}
%
%
\noindent
{\bf 1.\ Introduction.}
Recently, direct CP violation in $K\rightarrow\pi\pi$ decays was 
observed by the KTeV~\cite{ktev} and NA48~\cite{sozzi} collaborations. 
The new measurements are in agreement with the results of the NA31 
experiment~\cite{barr}. The present world average for the parameter 
$\varepsilon'/\varepsilon$ is~\cite{sozzi}
\begin{equation}
\mbox{Re}\,\varepsilon'/\varepsilon\,=\,(21.2 \pm 4.6)\cdot 10^{-4}\,,
\label{exp}
\end{equation}
which differs from zero by 4.6 standard deviations. In the standard 
model CP violation originates in the CKM phase, and direct CP violation 
is governed by loop diagrams of the penguin type, in which the three quark 
generations are present. The value of $\varepsilon'/\varepsilon$ is 
determined by a non-trivial interplay of the strong, electromagnetic,
and weak interactions and depends on almost all SM parameters. This makes 
its calculation quite complex. The main source of uncertainty in the 
calculation of the CP ratio is the QCD non-perturbative contribution 
related to the hadronic nature of the $K\rightarrow\pi\pi$ decays. 
Using the $\Delta S=1$ effective hamiltonian,
\begin{equation}
{\cal H}_{ef\hspace{-0.5mm}f}^{\SS\Delta S=1}=\frac{G_F}{\sqrt{2}}
\;\lambda_u\sum_{i=1}^8 c_i(\mu)\,Q_i(\mu)\hspace{1cm}(\mu < m_c)\,,
\label{ham}
\end{equation}
the non-perturbative contribution, contained in the hadronic matrix elements 
of the four-quark operators $Q_i$, can be separated, at the renormalization 
scale $\mu\simeq 1\,$GeV, from the perturbative Wilson coefficients 
$c_i(\mu)=z_i(\mu)+\tau y_i(\mu)$ (with $\tau=-\lambda_t/\lambda_u$ and 
$\lambda_q=V_{qs}^*\,V_{qd}^{}$). Introducing the matrix elements $\langle 
Q_i\rangle_I\equiv\langle(\pi\pi)_I|Q_i|K\rangle$, $\varepsilon'/\varepsilon$ 
can be written as 
\begin{equation}
\frac{\varepsilon'}{\varepsilon}\,=\,\frac{G_F}{2}
\frac{\omega\,\mbox{ Im}\lambda_t}{|\varepsilon|\,\mbox{Re}A_0}
\left[\,\Big|\sum_i\,y_i\,\langle Q_i\rangle_0\Big|\,
\Big(1-\Omega_{\eta+\eta'}\Big)\,\,-\,\frac{1}{\omega}
\Big|\sum_i\,y_i\,\langle Q_i\rangle_2\Big|\,\right].
\label{epspsm}
\end{equation}
$\omega=$Re$A_0/$Re$A_2=22.2$ is the ratio of the CP conserving $K
\rightarrow\pi\pi$ isospin amplitudes; $\Omega_{\eta+\eta'}$ takes 
into account the effect of the isospin breaking in the quark masses. 
In an exact realization of non-perturbative QCD, the scale dependences 
of the $y_i$ and $\langle Q_i\rangle_I$ in Eq.~(\ref{epspsm}) must 
cancel. $\varepsilon'/\varepsilon$ is dominated by $\langle Q_6
\rangle_0$ and $\langle Q_8\rangle_2$ which cannot be fixed from the 
CP conserving data, different from most of the matrix elements of the 
current-current operators \cite{BJM,bosch}. Beside the theoretical 
uncertainties coming from the calculation of the $\langle Q_i 
\rangle_I$ and of $\Omega_{\eta+\eta'}$, the analysis of the CP 
ratio suffers from the uncertainties on the values of various input 
parameters, in particular of the CKM phase in Im$\lambda_t$, of 
$\Lambda_{\mbox{\tiny QCD}}\equiv\Lambda^{(4)}_{\overline{\mbox{\tiny MS}}}$,
and of the strange quark mass. The uncertainty on $m_s$ is especially 
important due to the density-density structure of $Q_6$ and $Q_8$, which 
implies that their matrix elements are proportional to the square of the
quark condensate and are hence proportional to $1/m_s^2$. In all methods 
the calculation of the hadronic matrix element of $Q_6$, which is a 
peculiar operator, appears to be the most difficult one.

In this talk we present the results we obtained for $\varepsilon'/ 
\varepsilon$ together with G.O.\ K\"ohler and E.A.\ Paschos \cite{HKPS}. 
We briefly explain the method used to compute the hadronic matrix elements, 
and discuss in detail the peculiarities in the calculation of the matrix 
element of $Q_6$.

%
\vspace{0.3cm}
\noindent
{\bf 2.\ General Framework.}
To calculate the hadronic matrix elements we start from the effective chiral
lagrangian for pseudoscalar mesons which involves an expansion in momenta 
where terms up to ${\cal O}(p^4)$ are included \cite{GaL}. Keeping only 
terms of ${\cal O}(p^4)$ which contribute, at the order we calculate, to 
the $K \rightarrow \pi \pi$ amplitudes, for the lagrangian we obtain:
\begin{eqnarray}
{\cal L}_{ef\hspace{-0.5mm}f}&=&\frac{f^2}{4}\Big(
\langle D_\mu U^\dagger D^{\mu}U\rangle
+\frac{\alpha}{4N_c}\langle \ln U^\dagger -\ln U\rangle^2 
+\langle\chi U^\dagger+U\chi^\dagger\rangle\Big) \nonumber\\[1.5mm] 
&& 
+L_5\langle D_\mu U^\dagger D^\mu U(\chi^\dagger U+U^\dagger\chi)
\rangle +L_8\langle \chi^\dagger U\chi^\dagger U+\chi 
U^\dagger\chi U^\dagger \rangle\,,\hspace*{5mm}
\end{eqnarray}
with $\langle A\rangle$ denoting the trace of $A$, $\alpha=m_\eta^2
+m_{\eta'}^2-2m_K^2$, $\chi=r{\cal M}$, and ${\cal M}=\mbox{diag}
(m_u,m_d,m_s)$. $f$ and $r$ are parameters related to the pion decay 
constant $F_\pi$ and to the quark condensate, with $r=-2\langle\bar{q}q
\rangle/f^2$. The complex matrix $U$ is a non-linear representation of 
the pseudoscalar meson nonet. The conventions and definitions we use 
are the same as those in~\cite{HKPS,HKPSB,hks}. 

The method we use is the $1/N_c$ expansion introduced in~\cite{BBG2}. 
In this approach, we expand the matrix elements in powers of the external
momenta and of $1/N_c$. From the lagrangian the mesonic representations
of the quark currents and densities can be obtained by usual bosonization 
techniques. It is then straightforward to calculate the tree level 
(leading-$N_c$) matrix elements. For the $1/N_c$ corrections to the matrix 
elements $\langle Q_i\rangle_I$ we calculated chiral loops as described 
in~\cite{HKPSB,hks}. Especially important to this analysis are the 
non-factorizable $1/N_c$ corrections, which are UV divergent and must be 
matched to the short-distance part. They are regularized by a finite 
cutoff which is identified with the short-distance renormalization scale 
[8\,--\,11]. The definition of the momenta in the loop diagrams, which are 
not momentum translation invariant, is discussed in detail in~\cite{HKPSB}. 

For the Wilson coefficients we use both the leading logarithmic (LO) 
and the next-to-leading logarithmic (NLO) values~\cite{BJM}. In the 
pseudoscalar approximation, the matching has to be done below 1\,GeV. 
Values of the $y_i$ at scales $0.6\,\mbox{GeV}\leq\mu\leq 0.9\,\mbox{GeV}$ 
where communicated to us by M.~Jamin. The NLO values are scheme dependent 
and are calculated within naive dimensional regularization (NDR) and in 
the 't Hooft-Veltman scheme (HV). The absence of any reference to the 
renormalization scheme in the low-energy calculation, at this stage, 
prevents a complete matching at the next-to-leading order~\cite{ab98}. 
Nevertheless, a comparison of the numerical results obtained from the LO 
and NLO coefficients is useful as regards estimating the uncertainties 
and testing the validity of perturbation theory.

%
\vspace{0.3cm}
\noindent
{\bf 3.\ Analysis of \boldmath $\varepsilon'/\varepsilon$\unboldmath.}
At next-to-leading order, in the twofold expansion in powers of external 
momenta and of $1/N_c$, we investigate the tree level contributions from the 
${\cal O}(p^2)$ and the ${\cal O}(p^4)$ lagrangian as well as the one-loop 
contribution from the ${\cal O}(p^2)$ lagrangian. For the matrix elements 
of the current-current operators, the corresponding terms are ${\cal O}
(p^2)$, ${\cal O}(p^4)$, and ${\cal O}(p^2/N_c)$, respectively; for 
density-density operators they are ${\cal O}(p^0)$, ${\cal O}(p^2)$, 
and ${\cal O}(p^0/N_c)$. 

Analytical formulas for all matrix elements at these orders are given 
in~\cite{HKPSB,hks}. Among them four are particularly interesting and 
important:
\begin{eqnarray}
\langle Q_1\rangle_0
&=& -\frac{1}{\sqrt{3}}F_\pi\left( \mk^2 - \mp^2 \right)
\left[1+\frac{4\hat{L}_5^r}{F_\pi^2}m_\pi^2+
\frac{1}{(4\pi)^2F_\pi^2}\right.
\nonumber\\
&&\times\left.\left( 6\Lambda_c^2-\Big(\frac{1}{2}\mk^2
+6m_\pi^2\Big)\log\Lambda_c^2\,+\,\cdots\,
\right)\right]\,,\label{rm1}\\[1mm]
\langle Q_2\rangle_0
&=& \frac{2}{\sqrt{3}}F_\pi\left( \mk^2 - \mp^2 \right)
\left[1+\frac{4\hat{L}_5^r}{F_\pi^2}m_\pi^2+
\frac{1}{(4\pi)^2F_\pi^2}\right.
\nonumber \\
&&\left.\times\left(\frac{15}{4}\Lambda_c^2+\Big(\frac{11}{8}\mk^2
-\frac{15}{4}m_\pi^2\Big)\log\Lambda_c^2\,+\,\cdots\,\right)\right]\,,
\label{rm2}\\[1.5mm]
\langle Q_6\rangle_0
&=&-\frac{4\sqrt{3}}{F_\pi}R^2(m_K^2-m_\pi^2) 
\left[\hat{L}_5^r-\frac{3}{16\,(4\pi)^2}\,\log\Lambda_c^2
\,+\,\cdots\,\right]\label{rm3}\,,\\[1.5mm]
\langle Q_8\rangle_2
&=&\frac{\sqrt{3}}{2\sqrt{2}}F_\pi R^2
\left[1+\frac{8m_K^2}{F_\pi^2}\,\Big(\hat{L}_5^r-2\hat{L}_8^r\Big) 
-\frac{4m_\pi^2}{F_\pi^2}\,\Big(3\hat{L}_5^r-8\hat{L}_8^r\Big)\right.
\hspace*{4mm}\nonumber\\
&& 
-\left.\frac{1}{(4\pi)^2F_\pi^2}\Big(m_K^2-m_\pi^2+\frac{2}{3}
\alpha\Big)\log\Lambda_c^2 \,+\,\cdots\,\right]\,,\label{rm4}
\end{eqnarray}
with $R\equiv 2m_K^2/(m_s+m_d)$. The ellipses denote finite terms 
which are not written explicitly here. The constants $\hat{L_5^r}$ 
and $\hat{L_8^r}$ are renormalized couplings whose values are 
$\hat{L}_5^r=2.07\cdot 10^{-3}$ and $\hat{L}_8^r=1.09\cdot 10^{-3}$. 
Eqs.~(\ref{rm1})\,-\,(\ref{rm4}) have several interesting properties 
\cite{HKPSB,hks}. First, the VSA values for $\langle Q_1\rangle_0$ and 
$\langle Q_2\rangle_0$ are far too small to account for the large 
$\Delta I=1/2$ enhancement observed in the CP conserving amplitudes. 
Using the large-$N_c$ limit [$B_1^{(1/2)}=3.05$, $B_2^{(1/2)}=1.22$; 
$B_i^{(\Delta I)}\equiv\mbox{Re}\langle Q_i\rangle_I/\langle Q_i
\rangle_I^{\mbox{\tiny VSA}}$] improves the agreement between theory 
and experiment, but it still provides a gross underestimate. However, 
the non-factorizable $1/N_c$ corrections in Eqs.~(\ref{rm1}) 
and~(\ref{rm2}) contain quadratically divergent terms which are not 
suppressed with respect to the tree level contribution, since they bring 
about a factor of $\Delta \equiv\Lambda_c^2/(4\pi F_\pi)^2$ and have large 
prefactors. Varying $\Lambda_c$ between 600 and 900\,MeV, $B_1^{(1/2)}$ 
and $B_2^{(1/2)}$ take the values $8.2-14.2$ and $2.9-4.6$, respectively. 
Quadratic terms in $\langle Q_1\rangle_0$ and $\langle Q_2\rangle_0$ produce 
a large enhancement which brings the $\Delta I=1/2$ amplitude in agreement 
with the data \cite{hks}. Corrections beyond the chiral limit ($m_q=0$) in 
Eqs.~(\ref{rm1})\,-\,(\ref{rm2}) are suppressed by a factor of $\delta 
=m_{K,\pi}^2/(4\pi F_\pi)^2 \simeq 20\,\%$ and were found to be small. 

For the operators $Q_6$ and $Q_8$, values rather close to the VSA
[$B_6^{(1/2)}=B_8^{(3/2)}=1$] are used in the literature. As 
a result the experimental range for $\varepsilon'/\varepsilon$ can be 
accommodated in the standard model only if there is a conspiracy of the 
input parameters $m_s$, $\Omega_{\eta+\eta'}$, Im$\lambda_t$, and 
$\Lambda_{\mbox{\tiny QCD}}$ (see e.g.~\cite{bosch}). The fact that
the VSA fails completely in explaining the $\Delta I =1/2$ rule therefore
raises the question whether it can be used for $Q_6$ and $Q_8$. In 
fact, at the present stage of the calculation, the case of $\langle 
Q_6\rangle_0$ and $\langle Q_8\rangle_2$ is different from that of 
$\langle Q_{1,2}\rangle_0$. The leading-$N_c$ values are very close to 
the corresponding VSA values. Moreover, the non-factorizable loop 
corrections in Eqs.~(\ref{rm3}) and~(\ref{rm4}), which are of ${\cal O}
(p^0/N_c)$, are found to be only logarithmically divergent~\cite{HKPSB}. 
Consequently, in the case of $\langle Q_8\rangle_2$ they are suppressed by 
a factor of $\delta$ compared to the leading ${\cal O}(p^0)$ term and are 
expected to be of the order of $20\,\%$ to $50\,\%$ depending on the 
prefactors. We note that Eq.~(\ref{rm4}) is a full leading plus 
next-to-leading order analysis of the $Q_8$ matrix element. The case of 
$B_6^{(1/2)}$ is more complicated since the ${\cal O}(p^0)$ term vanishes 
for $Q_6$. Nevertheless, the non-factorizable loop corrections to this 
term remain and have to be matched to the short-distance part of the 
amplitudes \cite{HKPSB}. These ${\cal O}(p^0/N_c)$ non-factorizable 
corrections must be considered at the same level, in the twofold expansion, 
as the ${\cal O}(p^2)$ tree contribution. Consequently, a value of 
$B_6^{(1/2)}$ around one [which corresponds to the ${\cal O}(p^2)$ term 
alone] is not a priori expected. However, numerically it turns out that 
the ${\cal O}(p^0/N_c)$ contribution is only moderate. This property can 
be understood from the $(U^\dagger)_{dq}(U)_{qs}$ structure of the $Q_6$ 
operator which vanishes to ${\cal O}(p^0)$ implying that the factorizable 
and non-factorizable ${\cal O}(p^0/N_c)$ contributions cancel to a large 
extent~\cite{HKPSB}. This explains why the deviations we observe with 
respect to the VSA values are smaller than for $Q_1$ and $Q_2$ and why 
in particular for $Q_6$ to ${\cal O}(p^0/N_c)$ we do not observe a 
$\Delta I=1/2$ enhancement. Varying $\Lambda_c$ between 600 and
900\,MeV, $B_6^{(1/2)}$ and $B_8^{(3/2)}$ take the values $1.10-0.72$
and $0.64-0.42$, respectively. $B_6^{(1/2)}$ and $B_8^{(3/2)}$ are 
therefore more efficiently protected from possible large $1/N_c$ 
corrections of the ${\cal O}(p^2)$ lagrangian than $B_{1,2}^{(1/2)}$. 
The effect of the ${\cal O}(p^0/N_c)$ term is however important for 
$B_6^{(1/2)}$ as for $B_8^{(3/2)}$ because it gives rise to a noticeable 
dependence on the cutoff scale \cite{HKPSB}. We note that $B_8^{(3/2)}$ 
shows a scale dependence which is very similar to the one of $B_6^{(1/2)}$ 
leading to a stable ratio $B_6^{(1/2)}/B_8^{(3/2)}\simeq 1.72$ for 
$\Lambda_c$ between 600 and $900$\,MeV. The ${\cal O}(p^0/N_c)$ 
corrections consequently make the cancellation of $Q_6$ and $Q_8$ in 
$\varepsilon'/\varepsilon$ less effective, but the values of the matrix
elements are reduced. Hence we obtain values for $\varepsilon'/ 
\varepsilon$ \cite{HKPS} close to the ones found with the VSA values 
of $B_6^{(1/2)}$ and $B_8^{(3/2)}$; in particular, for central values of 
the input parameters [$m_s(1$~GeV)=150 MeV, $\Omega_{\eta+\eta'}=0.25$, 
Im$\lambda_t=1.33\cdot 10^{-4}$, and $\Lambda_{\mbox{\tiny QCD}}=325$\,MeV, 
see~\cite{HKPS} and references therein] the values for the CP ratio 
are significantly smaller than the data.

However, since the leading ${\cal O}(p^0)$ contribution vanishes for 
$Q_6$, corrections from higher order terms beyond the ${\cal O}(p^2)$ 
and ${\cal O}(p^0/N_c)$ are expected to be large \cite{HKPS}. The terms 
of ${\cal O}(p^2)$ and ${\cal O}(p^0/N_c)$ correspond to the lowest 
(non-vanishing) order, and the calculation of the next order terms 
is very desirable. In the twofold expansion, the higher order 
corrections to the matrix element of $Q_6$ are of orders: ${\cal O}(p^4)$, 
${\cal O}(p^0/N_c^2)$, and ${\cal O}(p^2/N_c)$. A full calculation of these 
terms is beyond the scope of our study. In particular, higher order terms 
in the $p^2$ expansion, which are chirally suppressed, cannot be calculated 
because the low-energy couplings in the ${\cal O}(p^6)$ lagrangian are very 
uncertain or even unknown. In~\cite{HKPS} we investigated the ${\cal O}
(p^2/N_c)$ contribution, i.e., the $1/N_c$ correction at the next order 
in the chiral expansion, because it brings about, for the first time, 
quadratic corrections on the cutoff. We remind the readers that for the 
CP conserving amplitude it is mainly the (quadratic) ${\cal O}(p^2/N_c)$ 
corrections which bring to $\langle Q_{1,2}\rangle_0$ a large enhancement 
relative to the (leading-$N_c$) ${\cal O}(p^2)$ values. As the leading-$N_c$ 
value for $Q_6$ is also ${\cal O}(p^2)$ we cannot a priori exclude that the 
value of $\langle Q_6\rangle_0$ is largely affected by ${\cal O}(p^2/N_c)$ 
corrections too, removing from $Q_6$ the property observed to ${\cal O}
(p^0/N_c)$, to be protected from large $1/N_c$ corrections. As explained 
above, quadratic ${\cal O}(p^2/N_c)$ corrections are proportional to the 
factor $\Delta\equiv\Lambda_c^2/(4 \pi F_\pi)^2$ relative to the 
${\cal O}(p^2)$ tree level contribution. Different is the case of $Q_8$ 
since its leading-$N_c$ value is ${\cal O}(p^0)$ at lowest order in the 
chiral expansion. Quadratic terms for $Q_8$ are consequently chirally 
suppressed with respect to the leading-$N_c$ value. 

Calculating the quadratic term of ${\cal O}(p^2/N_c)$ for matrix element
of $Q_6$ and adding it to the ${\cal O}(p^2)$ and ${\cal O}(p^0/N_c)$ 
result in Eq.~(\ref{rm3}) we obtain:
\begin{equation}
\hspace*{-1mm}
\langle Q_6\rangle_0
=-\frac{4\sqrt{3}}{F_\pi}R^2(m_K^2-m_\pi^2)\left[\hat{L}_5^r
\Bigg(1+\frac{3}{2}\frac{\Lambda_c^2}{(4\pi)^2 F_\pi^2}\Bigg)
-\frac{3}{16}\frac{\log\Lambda_c^2}{(4\pi)^2}+\cdots\,\right]
\hspace*{-1mm}.
\label{q6total}
\end{equation}
The result for the ${\cal O}(p^2/N_c)$ term we already presented in
\cite{PSOrsay}. Numerically, we observe a large positive correction from 
the quadratic term in Eq.~(\ref{q6total}). The slope of this correction is 
qualitatively consistent and welcome since it compensates for the logarithmic 
decrease at ${\cal O}(p^0/N_c)$. Varying $\Lambda_c$ between 600 and 
900\,MeV, the $B_6^{(1/2)}$ factor takes the values $1.50-1.62$. The 
approximate stability of $B_6^{(1/2)}$ is in accordance with the 
perturbative evolution, since the non-diagonal dependence of $y_6$ on 
the renormalization scale, i.e., the one beyond the leading-$N_c$ scale 
dependence of $R^2$ in Eq.~(\ref{q6total}), was found to be small. The 
quadratic term of ${\cal O}(p^2/N_c)$ is of the same magnitude as the 
${\cal O}(p^2)$ tree term. $Q_6$ is a $\Delta I=1/2$ operator, and the 
enhancement of $\langle Q_6\rangle_0$ indicates that at the level of the 
$1/N_c$ corrections the dynamics of the $\Delta I=1/2$ rule applies to 
$Q_6$ as to $Q_{1,2}$. One might however note that the enhancement 
observed for $Q_6$ is smaller than for $Q_1$ and $Q_2$.
 
Using the quoted values for $B_6^{(1/2)}$ together with the full 
leading plus next-to-leading order $B$ factors for the remaining 
operators \cite{HKPS} we calculated $\varepsilon'/\varepsilon$ for 
central values of $m_s$, $\Omega_{\eta+\eta'}$, Im$\lambda_t$, and 
$\Lambda_{\mbox{\tiny QCD}}$. The results for the three sets of Wilson 
coefficients LO, NDR, and HV and for $\Lambda_c$ between 600 and $900\,
\mbox{MeV}$ are given in Tab.~\ref{tab1}. The numbers are obtained with two 
different methods for analyzing the sensitivity on the imaginary part coming 
from the final states interactions. In the first case, we use the real part 
of our calculation and the phenomenological phases $\delta_0=(34.2\pm 
2.2)^\circ$ and $\delta_2=(-6.9 \pm 0.2)^\circ$ \cite{phases}, and 
replace $|\sum_i y_i\langle Q_i\rangle_I|$ in Eq.~(\ref{epspsm}) by 
$\sum_i y_i\mbox{Re}\langle Q_i\rangle_I/\cos\delta_I$ \cite{BEF}. In the 
second case, we use only the real part assuming zero phases. The latter
case is very close to the results we would get if we used the small 
imaginary part obtained at the one-loop level \cite{HKPS}. Collecting 
together the LO, NDR, and HV results for the two cases and for central 
values of the parameters, we find the following (conservative) range:
\begin{displaymath}
7.0\cdot 10^{-4}\,\,\leq\,\,\varepsilon'/\varepsilon
\,(\mbox{central})\,\,\leq\,\,24.7\cdot 10^{-4}\label{rangeeps}\,, 
\end{displaymath}
which is in the ball park of the experimental result in Eq.~(\ref{exp}). 
Performing a complete scanning of the parameters [$125\,\mbox{MeV}
\leq m_s(1\,\mbox{GeV})\leq 175\,\mbox{MeV}$, $0.15\leq\Omega_{\eta+\eta'}
\leq 0.35$, $1.04\cdot 10^{-4}\leq\mbox{Im}\lambda_t\leq 1.63\cdot 10^{-4}$, 
and $245\,\mbox{MeV}\leq\Lambda_{\mbox{\tiny QCD}}\leq 405\,\mbox{MeV}$] 
we obtain $2.2\cdot 10^{-4}\leq\varepsilon'/\varepsilon \,(\mbox{scanned})
\leq 63.2\cdot 10^{-4}$ (see Tab.~\ref{tab2}). The numerical values in the 
tables can be compared with the results of the Munich, Trieste, and 
Rome groups \cite{bosch,BEF,CM}. The values for $B_6^{(1/2)}$ can 
also be compared with \cite{BPdelta}. The large ranges reported in 
Tab.~\ref{tab2} can be traced back to the large ranges of the input 
parameters. The parameters, to a large extent, act multiplicatively, 
and the larger range for $\varepsilon'/\varepsilon$ is due to the fact 
that the central value(s) for the ratio are enhanced roughly by a factor 
of two compared to the results obtained with $B$ factors for $Q_6$ and 
$Q_8$ close to the VSA. More accurate information on the parameters, 
from theory and experiment, will restrict the values for the CP ratio.

%
\vspace{0.3cm}
\noindent
{\bf 4.\ Summary.}
We have shown that the operator $Q_6$, similar to $Q_1$ and $Q_2$, 
is not protected from large $1/N_c$ corrections coming from quadratic
terms of ${\cal O}(p^2/N_c)$. From general counting arguments we have good
indications that among the various next-to-leading order terms in the $p^2$
and $1/N_c$ expansions they are the dominant ones. Calculating those terms
we find that they enhance $B_6^{(1/2)}$ and bring $\varepsilon'/\varepsilon$
much closer to the data for central values of the parameters. We obtain a 
quadratic evolution for $Q_6$ which indicates that a $\Delta I=1/2$ 
enhancement is operative for $Q_6$ similar to the one of $Q_1$ and 
$Q_2$. $B_8^{(3/2)}$ is expected to be affected much less by terms of
${\cal O}(p^2/N_c)$ due to an extra $p^2$ suppression factor relative
to the leading ${\cal O}(p^0)$ tree term.
\noindent
\begin{table}[t]
\begin{eqnarray*}
\begin{array}{|c||c|c|}\hline
 & \mbox{Case}\,\, 1& \mbox{Case}\,\, 2\\
\hline\hline 
\rule{0cm}{5mm} 
\mbox{LO}& \,19.5 \,\,\leq\,\,\varepsilon'/\varepsilon\,\, 
\leq\,\,24.7 \, 
& \,14.8 \,\,\leq\,\,\varepsilon'/\varepsilon\,\, 
\leq\,\,19.4 \, \\[0.2mm]
\mbox{NDR}& \,16.1 \,\,\leq\,\,\varepsilon'/\varepsilon\,\, 
\leq\,\,23.4 \, 
& \,12.5 \,\,\leq\,\,\varepsilon'/\varepsilon\,\,
\leq\,\,18.3 \, \\[0.2mm]
\mbox{HV}& \,9.3\,\,\,\,\leq\,\,\varepsilon'/\varepsilon\,\, 
\leq\,\,19.3 \, 
& \,7.0\,\,\,\,\leq\,\,\varepsilon'/\varepsilon\,\,
\leq\,\,14.9 \, \\[0.2mm]
\hline
\,\mbox{LO}\,+\,\mbox{NDR}\,+\,\mbox{HV}\,
& \,9.3\,\,\,\,\leq\,\,\varepsilon'/\varepsilon\,\,\leq \,\,24.7\, 
& \,7.0 \,\,\,\,\leq\,\,\varepsilon'/\varepsilon\,\,
\leq\,\,19.4\,\\[0.4mm]
\hline
\end{array}
\end{eqnarray*}
\caption{Central ranges for $\varepsilon'/ \varepsilon$ (in units of
$10^{-4}$) as explained in the text.
\label{tab1}}
\end{table}
\begin{table}[t]
\begin{eqnarray*}
\begin{array}{|c||c|c|}\hline
 & \mbox{Case}\,\, 1& \mbox{Case}\,\, 2\\
\hline\hline 
\rule{0cm}{5mm} 
\mbox{LO}& \,8.0 \,\,\leq\,\,\varepsilon'/\varepsilon\,\, 
\leq\,\,62.1 \, 
& \, 6.1\,\,\leq\,\,\varepsilon'/\varepsilon\,\, 
\leq\,\, 48.5\, \\[0.2mm]
\mbox{NDR}& \,6.8 \,\,\leq\,\,\varepsilon'/\varepsilon\,\, 
\leq\,\,63.9 \, 
& \,5.2 \,\,\leq\,\,\varepsilon'/\varepsilon\,\,
\leq\,\,49.8 \, \\[0.2mm]
\mbox{HV}& \,2.8 \,\,\leq\,\,\varepsilon'/\varepsilon\,\, 
\leq\,\,49.8 \, 
& \,2.2 \,\,\leq\,\,\varepsilon'/\varepsilon\,\,
\leq\,\,38.5 \, \\[0.2mm]
\hline
\,\mbox{LO}\,+\,\mbox{NDR}\,+\,\mbox{HV}\,
& \,2.8 \,\,\leq\,\,\varepsilon'/\varepsilon\,\,\leq \,\,63.9\, 
& \,2.2 \,\,\leq\,\,\varepsilon'/\varepsilon\,\,
\leq\,\,49.8 \,\\[0.4mm]
\hline
\end{array}
\end{eqnarray*}
\caption{Same as in Tab.~\ref{tab1}, but for the complete
scanning of the parameters.
\label{tab2}}
\end{table}

One should recall that our analysis of the ${\cal O}(p^2/N_c)$ terms 
for $Q_6$ is performed in the chiral limit. It would be desirable to 
calculate the corrections beyond the chiral limit, from logarithms and 
finite terms. It would also be interesting to investigate the effect of 
higher resonances. Each of the additional effects separately is not 
expected to counteract largely the enhancement found for $B_6^{(1/2)}$. 
Nevertheless, in the extreme (and unlikely) case where all these effects 
would come with the same sign a significant modification of the result 
cannot be excluded formally. In order to reduce the scheme dependence 
in the result for $\varepsilon'/\varepsilon$, appropriate subtractions 
would be necessary (see~\cite{BPdelta,BB}).\\[5mm]
{\bf Acknowledgments}: The talk is based on the work carried through
with G.O.~K\"ohler and E.A.~Paschos \cite{HKPS}. This work was supported 
in part by BMBF, 057D093P(7), Bonn, FRG, and DFG Antrag PA-10-1. 
T.H.~acknowledges partial support from EEC, TMR-CT980169.
%
%

%
\renewcommand{\textfraction}{1.0}
\end{document}